\documentclass[preprint,12pt]{elsarticle}

\usepackage{amssymb}

\journal{Nuclear Physics B}

\newcommand{\gapprox}{\raisebox{-.2ex}{$\stackrel{\textstyle>}{\raisebox{-.6ex}[0ex][0ex]{$\sim$}}$}}

\newcommand{\bra}[1]{\langle{#1}|}
\newcommand{\braket}[1]{\langle{#1}\rangle}
\newcommand{\ket}[1]{|{#1}\rangle}
\newcommand{\rEdS}{\rho_{\mathrm{EdS}}}

\newcommand{\lto}{\raisebox{-.1ex}{$\stackrel{\textstyle<}{\raisebox{-.6ex}[0ex][0ex]{$\to$}}$}}
\newcommand{\degK}{^\circ\mathrm{K}~}
\newcommand{\mbo}[1]{$#1$}
\newcommand{\lpl}{\Lambda_{\rm Pl}}
\newcommand{\MPl}{M_{\rm Pl}}
\newcommand{\mpl}{M_{\rm Pl}}
\newcommand{\tpl}{t_{\rm Pl}}
\newcommand{\Tpl}{T_{\rm Pl}}

\newcommand{\power}[1]{\times 10^{#1}}
\newcommand{\crn}{\nn \\ }
\newcommand{\bea}{\begin{eqnarray}}
\newcommand{\eea}{\end{eqnarray}}
\newcommand{\Ba}{\begin{eqnarray}}
\newcommand{\Ea}{\end{eqnarray}}

\newcommand{\nn}{\nonumber}

\newcommand{\gv}{\mbox{GeV}}

\newcommand{\epo}{\,.}
\newcommand{\semis}{\,;\;\;}

\newcommand{\be}{\begin{equation}}
\newcommand{\ee}{\end{equation}}

\newcommand{\MSb}{$\overline{\rm MS}$ }

\newcommand{\E}{\mathrm{e}}

\newcommand{\np}{Nucl.\ Phys.\ B }
\newcommand{\prl}{Phys.\ Rev.\ Lett.\ }

\begin{document}

\begin{frontmatter}



\title{%
\vskip-3cm{\baselineskip14pt
\centerline{\normalsize DESY~15-031,~~HU-EP-15/10\hfill}
}
\vskip1.5cm
The hierarchy problem and the cosmological constant problem in the Standard Model}


\author{Fred Jegerlehner}

\address{Humboldt-Universit\"at zu Berlin, Institut f\"ur Physik,
       Newtonstrasse 15, D-12489 Berlin, Germany\\
Deutsches Elektronen-Synchrotron (DESY), Platanenallee 6, D-15738 Zeuthen, Germany}

\begin{abstract}
We argue that the SM in the Higgs phase does not suffer form a
``hierarchy problem'' and that similarly the ``cosmological constant
problem'' resolves itself if we understand the SM as a low energy
effective theory emerging from a cut-off medium at the Planck scale.
We discuss these issues under the condition of a stable Higgs vacuum,
which allows to extend the SM up to the Planck length. The bare Higgs
boson mass then changes sign below the Planck scale, such the the SM
in the early universe is in the symmetric phase. The cut-off enhanced
Higgs mass term as well as the quartically enhanced cosmological
constant term trigger the inflation of the early universe. Reheating
follows by the heavy Higgses decaying predominantly into top--anti-top
pairs, which at this stage are sill massless. Preheating is suppressed
in SM inflation as in the symmetric phase bosonic decay channels are
absent at tree level. The coefficients of the shift between bare and
renormalized Higgs mass $m^2_{H 0}-m^2_{H}=\delta
m^2=\frac{\lpl^2}{16\pi^2}\,C(\mu)$ as well as of the shift between
bare and renormalized vacuum energy density
$\rho_{\Lambda 0}-\rho_{\Lambda}=\delta \rho_{\rm vac}
=\frac{\lpl^4}{(16\pi^2)^2}\,X(\mu)$ exhibit close-by zeros $C(\mu)=0$ at
about $\mu_0\approx 1.4 \times 10^{16}~\gv$ or $\mu'_0\approx 7.7
\times 10^{14}~\gv$ after Wick rearrangement $C(\mu)\to
C'(\mu)=C(\mu)+\lambda(\mu)$ and
$X(\mu)=\frac18\,(2\,C(\mu)+\lambda(\mu))=0$ at $\mu_{\rm CC}\approx
3.1\times 10^{15}~\gv$. The zero of $C(\mu)$ triggers the electroweak
phase transition with $C(\mu) <0$ in the low energy Higgs phase and
$C(\mu)>0$ in the symmetric phase above the transition point. Since
inflation tunes the total energy density to take the critical value of
a flat universe $\Omega_{\rm tot}=\rho_{\rm tot}/\rho_{\rm crit}
=\Omega_\Lambda+\Omega_{\rm matter}+\Omega_{\rm radiation}=1$ it is
obvious that $\Omega_\Lambda$ today is of order $\Omega_{\rm tot}$
given that $1>\Omega_{\rm matter},\Omega_{\rm
radiation}>0$ which saturate the total density to about 26\% only, the
dominant part being dark matter(21\%). Obviously, the SM Higgs system
initially provides a huge \textit{dark energy} density and the
resulting inflation is taming the originally huge cosmological
constant to the small value observed today, whatever its initial value was,
provided it was large enough to trigger inflation. While laboratory
experiments can access physics of the broken phase only, the symmetric
phase above the Higgs transition point is accessible though physics of
the early universe as it manifests in cosmological observations. The
main unsolved problem in our context remains the origin of dark matter.
\end{abstract}


\end{frontmatter}


\section{Introduction}
The discovery of the Higgs boson~\cite{Englert:1964et,Higgs:1964ia} by
the ATLAS~\cite{ATLAS} and CMS~\cite{CMS} experiments at CERN revealed
a very peculiar value for the Higgs boson mass, just in a very narrow
window which allows to extrapolate the SM way up to the Planck
scale~\cite{Hambye:1996wb}. ATLAS and CMS results therefore may
``revolution'' particle physics in an unexpected way, namely showing
that the SM has higher self-consistency (conspiracy) than expected and
previous arguments for the existence of new physics may turn out not
to be compelling. Also the absence so far of any new physics signal at
the LHC may indicate that commonly accepted expectations may not be
satisfied. On the one hand it seems to look completely implausible to
assume the SM to be essentially valid up to Planck energies, on the
other hand the high tide of speculations about physics beyond the SM
have been of no avail. One also has to keep in mind that precision
tests of the SM already revealed a test in depth of its quantum
structure, besides large corrections form the running fine structure
constant $\alpha(s)$, the running of the strong coupling $\alpha_s(s)$
and the large top Yukawa $y^2_t(s)$ effect as contributing to the
$\rho=G_{\rm NC}/G_\mu(0)$ parameter, subleading corrections amount to
a 10 $\sigma$ deviation form the SM leading order effects
predictions. Thus the SM is on very solid grounds better than
everything else we ever had.

On the other hand the view that the SM is a low energy effective
theory of some cutoff system at the Planck energy scale $\MPl$ appears
to be consolidated. A crucial point is that $\MPl$ providing the scale
for the low energy expansion in powers $E/\MPl$ is exceedingly high,
very far from what we can see! A dimension 6 operator at LHC energies
is suppressed by $(E_{\rm LHC}/\lpl)^2\approx 10^{-30}\,$. This seems
to motivate a change in paradigm from the view that the world looks
simpler the higher the energy to a more natural scenario which
understands the SM as the ``true world'' seen from far away, with
symmetries emerging from not resolving the details.

The methodological approach for constructing low energy effective
theories we have learned from Ken
Wilson's~\cite{Wilson:1971bg,Jegerlehner:1976xd,Jegerlehner:2013cta}
investigations of condensed matter systems and his insight that
critical long distance phenomena are governed by emergent quantum
field theories. As I will argue in the following, cut-offs in particle
physics are important to understand early
cosmology~\cite{Jegerlehner:2013cta}, such as inflation, reheating,
baryogenesis and all that~\cite{Starobinsky:1980te,Guth:1980zm,Linde:1981mu,Kolb:1990vq,Weinberg:2008zzc}. As
in condensed matter physics the connection between macroscopic long
distance physics (at laboratory scales) and the microscopic underlying
cut-off system (high energy events as they were natural in the early
universe) turn out to have a physical meaning.

In this context naturalness
arguments play an important role. The SM's naturalness problems and
fine-tuning problems have been made conscious by G. 't
Hooft~\cite{'tHooft:1979bh} long time ago as a possible problem in the
relationship between macroscopic phenomena which follow from
microscopic laws (a condensed matter system inspired scenario), soon later the
``hierarchy problem'' had been dogmatized as a kind of fundamental
principle. In fact the hierarchy problem of the SM seems to be the key
motivation for all kind of extensions of the SM. It is therefore
important to reconsider the ``problem'' in more detail.

One of my key points concerns the different meaning a possible
hierarchy problem has in the symmetric and in the broken phase of the
SM.  In order to understand the point we have to remember why we need
the Higgs in the SM. The Higgs is necessary to get a renormalizable
low energy effective electroweak
theory~\cite{Glashow61}. Interestingly, one scalar particle is
sufficient to solve the renormalizability problems arising form each
of many different massive fields in the SM, of which each causes the
problem independently of the others. The point is that this one
particle has to exhibit as many new forces as there are individual
massive states~\cite{Weinberg67}. All required new interactions are in
accordance with the SM symmetry structure in the symmetric phase as we
know. The taming of the high energy behavior of course requires the
Higgs boson to have a mass in the ballpark of the other given heavier
SM states, if it would be much heavier it would not serve its purpose
in the low energy regime. Note that the Higgs boson has to cure the
unphysical mass effects for the \textbf{given} gauge boson masses
$M_W$, $M_Z$ and fermion masses $M_f$\footnote{We denote on-shell
masses by capital, \MSb masses by lower case letters as in Ref.~\cite{Jegerlehner:2013cta}}, via adequate \textit{Higgs
exchange forces}, where the coupling strength is proportional to the
mass of the massive field coupled. A very heavy Higgs eventually would
decouple and thus miss to restore renormalizability of the massive
vector-boson gauge theory. Interestingly, in the symmetric phase the
SM gauge-boson plus chiral fermions sector is renormalizable without
the Higgs-boson and Yukawa sectors and scalars are not required at all
to cure the high energy behavior, because it is renormalizable on its
own structure. Therefore, in the symmetric phase the mass-degenerate
Higgs fields in the complex Higgs doublet can be as heavy as we
like. Since unprotected by any symmetry, naturally we would expect the
Higgses indeed to be very heavy. Indeed, the ``origin'' of the Higgs
mass is very different in the broken phase, where the mass is
generated by the Higgs mechanism~\cite{Englert:1964et,Higgs:1964ia}
also for the Higgs itself ($m^2_H=\frac13\,\lambda\,v^2$), and in the
symmetric phase, where is is dynamically generated by the Planck
medium, as we will argue below. Therefore, the usual claim that the SM
requires to be extended in such a way that quadratic divergences are
absent has no foundation. Purely formal arguments based on
perturbative counterterm adjustments do not lead any further.
 
The hierarchy problem in particular addresses the presence of
quadratic ultraviolet (UV) divergences related with the SM Higgs mass
term. Infinities in physical theories are the result of idealizations
and show up as singularities in a formalism or in models. UV
singularities in general plague the precise definition as well as
concrete calculations in quantum field theories (QFT). A closer look
usually reveals infinities to parametrize our ignorance or mark the
limitations of our understanding or knowledge. One particular
consequence of UV divergences in local QFTs is that a vacuum energy is
ill-defined as it is associated with quartically divergent quantum
fluctuations.

This is another indication which tells us that local continuum QFT has
its limitation and that the need for regularization is actually the
need to look at the true system behind it. In fact the cut-off system is
more physical and does not share the problems with infinities which
result from the idealization. In any case the framework of a
renormalizable QFT, which has been extremely successful in particle
physics up to highest accessible energies, is not able to give answers
to the questions related to vacuum energy and hence to all questions
related to dark energy, accelerated expansion and inflation of the
universe. 

It is thus natural to consider the Standard Model to be what we
observe as the low energy effective SM (LEESM), the renormalizable
tail of the real cutoff system sitting at the Planck scale. As a
consequence all properties required by renormalizability, gauge
symmetries, chiral symmetry, anomaly cancellation naturally emerge as
a consequence of the low energy expansion. The infinite tower of
higher order operators becomes invisible, and only a few operators are
effectively observable, which makes the world look much simpler. In
reality infinities are replaced by eventually very large but finite
numbers, and I will show that sometimes such huge effects are needed
to understand the real world. I will argue that cutoff enhanced
effects are responsible for triggering the Higgs mechanism not very
far below the Planck scale and the inflation of the early universe.

The Planck medium, \textit{the ether}, is characterized by a fundamental cutoff
\mbo{\lpl} or equivalently the Planck mass \mbo{\mpl} 
which derive from the basic fundamental constants, the speed of light
\mbo{c} characterizing special relativity, the Planck constant
\mbo{\hbar} intrinsic to quantum physics and Newton's constant
\mbo{G_N} the key parameter of gravity. \textbf{Unified} they provide an
intrinsic length \mbo{\ell_\mathrm{Pl}}, the Planck length, which also
translates into the Planck time \mbo{\tpl} and the Planck temperature
\mbo{\Tpl}.

The history of our universe we can trace back 13.7 billion years close
to the Big Bang, when the expansion of the universe was ignited in a
``fireball'', an extremely hot and dense state when all structures and
at the end all atoms, nuclei and nucleons were disintegrated to a
world of elementary particles only. Besides the missing cold dark
matter (DM), one of the last piece which was missing in the SM, the
Higgs boson, now seems to provide a new milestone in our understanding
of the dynamics of the very early universe.

I think questions concerning the early universe can be addressed only
in the LEESM ``extension'' of the SM as such, given by a local QFT
supplied by cutoff effects in a minimal way. As we know, in a
renormalizable QFT all renormalized quantities as a function of the
renormalized parameters and fields in the limit of a large cut-off are
finite and devoid of any cut-off relicts! Here, it is adequate to
remember the Bogoliubov-Parasyuk renormalization theorem which states
that renormalized Green's functions and matrix elements of the
scattering matrix (S-matrix) are free of ultraviolet divergences. It
implies that in the low energy world cut-off effects are not
accessible to experiments and a ``problem'' like the hierarchy
problem is not a statement which can be checked to exist as a
observable conflict.

To my knowledge the only non-perturbative definition of a
renormalizable local quantum field theory is the possibility to put in
on a lattice. This again may be taken as an indication that the need
for a cut-off actually is an indication that the cutoff exists in the
real(er) world. In this sense the lattice QFT is the true(er) system than
its continuum tail. Of course, there are many ways to introduce a
cut-off and actually we cannot know what the cutoff system looks
truly. This is not a real problem if we are interested in the long
range patterns mainly, the only thing we have to care is that the
underlying system is in the \textit{universality class} of the SM.

\section{The hierarchy problem revisited}
The hierarchy problem cannot be
addressed within the renormalizable and renormalized SM, which is what we
can confront with experiments. In this framework all independent
parameters are free and have to be supplied by experiment.

In the LEESM ``extension'' of the SM bare parameter turn into physical
parameters of the underlying cut-off system as the ``true world'' at
short distances. Then the hierarchy problem is the problem of
``tuning to criticality'' which concerns the relevant 
operators of dimension $<4$, in particular the mass terms.
In the symmetric phase of the SM, where there is
only one mass (the others are forbidden by the known chiral and gauge
symmetries), the one in the potential of the Higgs doublet field,
the fine tuning to criticality has the form
\bea
m_0^2(\mu_1=\mpl)=m^2(\mu_2=M_H)+\delta m^2(\mu_1,\mu_2)\;;\;\; \delta m^2=
\frac{\lpl^2}{16 \pi^2}\,C(\mu)
\label{massren}
\eea
with a coefficient typically $C=O(1)$. To keep the renormalized mass $m$
at some small value, which can be seen at low energy, the bare $m^2_0$ has to
be adjusted to compensate the huge number $\delta m^2$ such that about
35 digits must be adjusted in order to get the observed value around
the electroweak scale. Is this a real problem?

One thing is apparent: our fine-tuning relation exhibits quantities at
very different scales, the renormalized one at low energy and the bare
one at the Planck scale. In the LEESM both are observable, in
principle. In fact, if we consider a renormalization condition like
(\ref{massren}), our presumed fine-tuning relation, if we want to test
it experimentally, it is not possible to test it by low energy
experiments only. Low energy experiments only allow us to test
relations between measurable renormalized quantities. While in a
renormalizable theory, relations between measurable quantities are
devoid of any cutoff effects, this changes when we perform high energy
experiments at a scale sensitive to the cutoff. Although such
experiments are not possible with down to earth accelerators, the
expansion of the universe has provided us a scan from Planck energies
down the \mbo{3~\degK} of the Cosmic Microwave Background
(CMB)~\cite{Ade:2013ktc} radiation. Thus, in the early universe a
relation like (\ref{massren}) has a direct physical meaning. In the SM
at low energies we are in the broken phase, where $m_{H 0}\approx
\delta m_{H}$ is huge negative, when looked at 
in a formal perturbative bookkeeping. However, this is not something we can
test by observation. If we want to test it we have to go to the short
distance scale, which however automatically flips the sign of $\delta
m^2_H$ and we automatically end up in the symmetric phase, where the
relation gets a different meaning. The key observation is that the
running SM parameters conspire in such a way that the Higgs mass
counterterm as well as the vacuum energy counterterm exhibit a zero,
which provides a matching point between the bare short distance world
and the renormalized low energy world. While the low energy world is
what we access by laboratory experiments, the high energy world is
what has shaped the early universe with the observable consequences we
see in cosmology, specifically in the CMB, Baryogenesis,
Nucleosynthesis and structure formation. We thus are able to learn
more about the short distance structure by making use of the early
universe as a natural accelerator.

In the Higgs phase, there is no hierarchy
problem~\cite{Jegerlehner:2013nna} (see also ~\cite{Malinsky:2012tp}).
It is true that in the relation $m^2_{H 0}=m^2_{H}+\delta m_H^2$ both
$m^2_{H 0}$ and $\delta m^2_{H}$ are many many orders of magnitude
larger than $ m^2_{H}\,.$ However, in the broken phase one
automatically obtains $ m^2_{H}\propto v^2(\mu_0)$, which is $O(v^2)$
not $ O(\MPl^2)$, irrespective of what the, at this scale
unobservable, objects of the bare theory are. Thus, in the broken
phase the Higgs is naturally light. That the Higgs mass likely is $
O(\MPl)$ in the symmetric phase is what promotes the Higgs to a
candidate for the inflaton.

One indeed can avoid artificial large numbers to show up by choosing $\mu_0$ 
as a renormalization point, where $\delta m^2=0$ and $m^2(\mu_0)=m^2_0(\mu_0)$ and
after the EW phase transition and corresponding vacuum rearrangement
$m^2_{H}(\mu_0)=2m^2(\mu_0)=\frac13\lambda(\mu_0) v^2(\mu_0)$ and then
get the physical Higgs mass by standard RG running and
matching. Certainly not the most practical way to implement $M_H$ as a
physical input parameter. The point is that in principle it is
possible circumvent fine-tuning. We also note that unlike in
regularized renormalizable QFT thinking, $m^2_0$ is not a given basic
parameter to be adjusted by renormalization. In the LEESM the bare
mass $m^2_0(\mu)$ as an effective mass, dynamically generated in the
Planck medium, is obviously also a running mass. Not far below the scale
$\mu_0$ the universe undergoes the EW phase transition (a point of
no-analyticity) and the Higgs mass is generated by a different
mechanism: the Higgs mechanism, the Higgs mass being given now by
$m^2_H=\frac13\,\lambda\,v^2$ after vacuum
rearrangement~\cite{Jegerlehner:2014mua}.  In the broken phase the
hierarchy problem is a pseudo problem.

In the broken phase, characterized by the
non-vanishing Higgs field vacuum expectation value (VEV) $v(\mu) \neq 0$,
all the masses are determined by the well known mass-coupling
relations 
\begin{eqnarray}
 m_W^2(\mu^2)&=&\frac14\,g^2(\mu^2)\,v^2(\mu^2)\semis
m_Z^2(\mu^2)=\frac14\,(g^2(\mu^2)+g'^2(\mu^2))\,v^2(\mu^2)\semis\crn
m_f^2(\mu^2)&=&\frac12\,y^2_f(\mu^2)\,v^2(\mu^2)\semis
m_H^2(\mu^2)=\frac13\,\lambda(\mu^2)\,v^2(\mu^2)\epo
\label{masscoupl}
\end{eqnarray}
Here we consider the parameters in the \MSb renormalization scheme,
$\mu$ is the \MSb renormalization scale, which we have to identify with
the energy scale of the physical processes or equivalently with the
corresponding temperature in the evolution of the universe.  The RG
equation for $v^2(\mu^2)$ follows from the RG equations for masses and
massless coupling constants using one of these relations.  As a key
relation we
use~\cite{Jegerlehner:2001fb}
\begin{eqnarray}
\mu^2 \frac{d}{d \mu^2} v^2(\mu^2)
=3\, \mu^2 \frac{d}{d \mu^2} \left[\frac{m_H^2(\mu^2)}{\lambda(\mu^2)} \right]
\equiv
v^2(\mu^2) \left[\gamma_{m^2}  - \frac{\beta_\lambda}{\lambda} \right]\,,
\label{vev}
\end{eqnarray}
where $\gamma_{m^2} \equiv \mu^2 \frac{d}{d \mu^2} \ln m^2$ and
$\beta_\lambda \equiv \mu^2 \frac{d}{d \mu^2} \lambda \,.$ We write
the Higgs potential as $V=\frac{m^2}{2}\,H^2+\frac{\lambda}{24}H^4$,
which fixes our normalization of the Higgs self-coupling. When the
$m^2$-term changes sign and $\lambda$ stays positive, we know we have
a first order phase transition.
Funny enough, the Higgs get its mass from its interaction with its own
condensate! and thus gets masses in the same way and in the same
ballpark as the other SM species. As mentioned before the Higgs mass
cannot by much heavier than the other heavier particles if
renormalizability is to be effective at low and moderate energies. The
interrelations (\ref{masscoupl}) also show that for fixed $v$, as determined by
the Fermi constant $G_\mu=1/(\sqrt{2}\,v^2)$, the Higgs cannot get too heavy if perturbation
theory should remain applicable. An extreme point of view claims that
naturalness requires all particles to have masses \mbo{O(\mpl)}
i.e. \mbo{v=O(\mpl)}. This would mean that the symmetry is not
restored at the cutoff scale and the notion of spontaneous symmetry
breaking (SSB) would be obsolete as a concept! The SM's successful
structure relies on a symmetric Lagrangian, and a ground state which
breaks the symmetry of the Lagrangian. The ground state is not
residing at the cutoff scale and breaks the symmetry in such a way
that the UV structure is not affected. In view that $v
\equiv 0$ above the EW phase transition point, why should it be
natural to expect that $v$ jumps from $0$ to $O(\mpl)$ during the
phase transition. We also note that as $v\to0$ all masses vanish,
with the exception of the Higgs mass which acquires a large value in
the symmetric phase (see below).

The Higgs VEV \mbo{v} is an \textbf{order parameter} resulting form
long range \textbf{collective behavior} and can be as small as we
like. Prototype is the magnetization $M$ as a function of temperature
$T$ in a ferromagnetic spin system, where \mbo{M=M(T)} and actually
\mbo{M(T)\equiv 0} for \mbo{T>T_c} and furthermore \mbo{M(T) \to 0} as
\mbo{T\lto T_c}.  For a direct non-perturbative check in case of the
SM, one would put the SM in the unitary gauge on a lattice and
simulate its long range properties. The Higgs boson VEV is then a well
defined physical order parameter. Difficulties related to Elitzur's
theorem~\cite{Elitzur:1975im} thus can be avoided.

Small \mbo{v/\mpl\ll 1} just means we are close to just below a
2$^\mathrm{nd}$ order phase transition point, which is not unnatural
if we take into consideration that long range behavior of condensed
matter systems are effective quantum field theories in a vicinity of
second order phase transition points~\cite{Wilson:1971bg}.

In the mass renormalization relation (\ref{massren}) the renormalized
mass measures the distance from the critical bare mass $m_{0c}$ for
which the renormalized $m$ is zero: thus $m^2=m_0^2-m^2_{0c}$. A
particle is seen at low energy only if it is light. In the symmetric
phase (short distance regime) is is natural to have both $m_0^2$ and
$\delta m^2$ large, but why $m_0^2 \approx \delta m^2$ to such
high precision?

At very high energy we see the bare system and the
Higgs field is a \textbf{collective field} which acquires its effective
mass via radiative effects $m^2_0 \approx
\delta m^2$ near below $\mpl$. In particle physics a
radiatively induced mass in known from the Coleman-Weinberg
mechanism~\cite{Coleman:1973jx}, now in the symmetric phase and
applied to the Planck medium. Such mechanism, which is natural in this
context, eliminates a possible fine-tuning problem at all
scales. There are many examples in condensed matter systems, like the
effective mass of the photon in the superconducting phase (Meissner
effect) or the effective mass of the effective field which encodes the
spin-singlet electron (Cooper) pairs in the Ginzburg-Landau
model~\cite{GLtheory} of superconductivity. The latter directly
corresponds to the Abelian Higgs model.

\section{Running SM parameters trigger the Higgs mechanism}
We remind that all dimensionless couplings satisfy the same
renormalization group (RG) equations in the broken and in the unbroken
phase and are not affected by any power cutoff dependencies. The
evolution of SM couplings in the \MSb scheme up to the Planck scale
has been investigated in
Refs.~\cite{Hambye:1996wb,Holthausen:2011aa,Yukawa:3,degrassi,Moch12,Mihaila:2012fm,Chetyrkin:2012rz,Masina:2012tz,
Bednyakov:2012rb,Tang:2013bz,Buttazzo:2013uya}
recently, and has been extended to include the Higgs VEV and the
masses in Refs.~\cite{Jegerlehner:2012kn,Jegerlehner:2013cta}. Except
for $g'$, which increases very moderately, all other couplings
decrease and stay positive up to the Planck scale. This strengthens
the reliability of perturbative arguments and reveals a stable Higgs
potential up to the Planck
scale~\cite{Jegerlehner:2012kn,Jegerlehner:2013cta}. While most
analyses~\cite{Yukawa:3,degrassi,Moch12,Masina:2012tz,Buttazzo:2013uya}
are predicting that for the given Higgs mass value vacuum stability is
nearby only (meta-stability), and actually fails to persist up to the
Planck scale, our evaluation of the matching conditions yields initial
\MSb parameters at the $Z$ boson mass scale which evolve preserving
the positivity of $\lambda$. Thereby the critical parameter is the top
quark Yukawa coupling, for which we find a slightly lower value. My
\MSb input at $M_Z$ is~\cite{Jegerlehner:2013cta} $g_3=1.2200$,
$g_2=0.6530$, $g_1=0.3497$, $y_t=0.9347$ and $\lambda=0.8070$. At
\mbo{\mpl} I get $g_3=0.4886$, $g_2=0.5068$, $g_1=0.4589$,
$y_t=0.3510$ and $\lambda=0.1405$.  In view of the fact that the
precise meaning of the experimentally extracted value of the top quark
mass is not free of ambiguities, usually it is identified with the
on-shell mass $M_t$ (see e.g.~\cite{Jegerlehner:2012kn} and references
therein), it may be premature to claim that instability of the SM
Higgs potential is a proven fact already. I also think that the
implementation of the matching conditions is not free of ambiguities,
while the evolution of the couplings over many orders of magnitude is
rather sensitive to the precise values of the initial
couplings. Accordingly, all numbers presented in this article depend
on the specific input parameters adopted, as specified in
Ref.~\cite{Jegerlehner:2012kn,Jegerlehner:2013cta}. In case the Higgs
self-coupling has a zero $\lambda(\mu^2)=0$, at some critical scale
$\mu_c$ below $\MPl$, we learn from Eq.~(\ref{vev}), or more directly
from $v(\mu^2)=\sqrt{6m^2(\mu^2)/\lambda(\mu^2)}\stackrel{\lambda \to
+0}{\to} \infty$ that the SM looses it meaning above this singular
point oft non-analyticity.

Running couplings can affect dramatically the quadratic divergences and the
interpretation of the hierarchy problem.
Quadratic divergences have been investigated at one loop in
Ref.~\cite{Veltman:1980mj} (see
also~\cite{Decker:1979cw,Degrassi:1992ff,Fang:1996cn}), at two loops
in Refs.~\cite{Alsarhi:1991ji,Hamada:2012bp,Jones:2013aua}. 
At $n$ loops the quadratic cutoff dependence is of the form
\be
\delta m_H^2= \frac{\Lambda^2}{16\pi^2}\,C_n(\mu)
\label{quadraic1}
\ee
where the n-loop coefficient only depends on the gauge couplings $g'$,
$g$, $g_3$, the Yukawa couplings $y_f$ and the Higgs self-coupling
$\lambda$. Neglecting the numerically insignificant light fermion
contributions, the one-loop coefficient function $C_1$ may be written
as
\bea
C_1=2\,\lambda+\frac32\, {g'}^{2}+\frac92\,g^2-12\,y_t^2
\label{coefC1}
\eea
and is uniquely determined by dimensionless couplings. The latter are
not affected by quadratic divergences such that standard RG equations
apply. Surprisingly, as first pointed out in
Ref.~\cite{Hamada:2012bp}, taking into account the running of the SM
couplings, the coefficient of the quadratic divergences of the bare
Higgs mass correction can vanish at some scale. In contrast to our
evaluation Hamada et al. actually find the zero to lie above the
Planck scale. In our analysis, relying on matching conditions for the
top quark mass analyzed in~\cite{Jegerlehner:2012kn}, we get a
scenario where $\lambda(\mu^2)$ stays positive up to the Plank scale
and looking at the relation between the bare and the renormalized
Higgs mass we find $C_1$ and hence the Higgs mass counterterm to
vanish at about $\mu_0\sim 1.4 \times 10^{16}~\gv$, not very far below
the Planck scale. The next-order correction, first calculated in
Refs.~\cite{Alsarhi:1991ji,Jones:2013aua} and confirmed
in~\cite{Hamada:2012bp} read
\bea
C_2&=&C_1+ \frac{\ln (2^6/3^3)}{16\pi^2}\, [
18\,y_t^4+y_t^2\,(-\frac{7}{6}\,{g'}^2+\frac{9}{2}\,g^2
             -32\,g_s^2) \nn \\
             &&-\frac{87}{8}\,{g'}^4-\frac{63}{8}\,g^4 -\frac{15}{4}\,g^2{g'}^2
             +\lambda\,(-6\,y_t^2+{g'}^2+3\,g^2)
             -\frac{2}{3}\,\lambda^2]\,,
\label{coefC2}
\eea
numerically does not change significantly the one-loop result.  The
same results apply for the Higgs potential parameter $m^2$, which
corresponds to $m^2\hat{=}\frac12\,m_H^2$ in the broken phase. For
scales $\mu <
\mu_0$ we have $\delta m^2$ large negative, which is triggering
spontaneous symmetry breaking by a negative bare mass
$m_0^2=m^2+\delta m^2$, where $m$ again denotes the renormalized mass. At
$\mu=\mu_0$ we have $\delta m^2=0$ and the sign of $\delta m^2$ flips,
implying a phase transition to the symmetric phase. Finite temperature
effects, which must be included in a realistic scenario, turn out not
do to change the gross features of our
scenario~\cite{Jegerlehner:2013cta}.
\section{The SM cosmological constant and dark energy}
It is crucial that in the early universe both terms in the Higgs
potential are positive in order to condition slow-roll inflation during long enough time.
In fact the quadratically and quartically cutoff enhanced terms in the
Higgs potential enforce the condition $\frac12\,\dot{\phi}^2 \ll
V(\phi)$ and given the Higgs boson pressure $p_\phi=\frac12\,\dot{\phi}^2-
V(\phi)$ and the Higgs energy density $\rho_\phi=\frac12\,\dot{\phi}^2
+ V(\phi)$, we arrive at the equation of state $w=p/\rho\approx -1$
  characteristic for dark energy and the equivalent cosmological constant (CC) 
(see e.g.~\cite{Straumann:1999ia,Volovik:2005zu,Sola:2013gha,Bass:2014lja}
and references therein).  A first measurement of the dark energy
equation of state $w=-1.13^{+0.13}_{-0.10}$, has been obtained by the
Planck mission~\cite{Ade:2013zuv} recently.

A key point is that in the LEESM scenario the vacuum energy
is a calculable quantity. In the symmetric phase $SU(2)$ symmetry
implies that \mbo{\Phi^+\Phi} is a singlet such that the invariant
vacuum energy is given just by simple Higgs loops~\cite{Jegerlehner:2014mua}
\bea
\bra{0}\Phi^+\Phi\ket{0}=\frac12\bra{0}H^2\ket{0}\equiv
\frac12\,\Xi\semis\Xi=\frac{\lpl^2}{16\pi^2}\,.
\label{vacenergy}
\eea 

A Wick type of rearrangement of the Lagrangian then provides a CC
represented by
$V(0)=\braket{V(\phi)}=\frac{m^2}{2}\,\Xi+\frac{\lambda}{8}\,\Xi^2$
and a mass shift ${m'}^{2}=m^2+\frac{\lambda}{2}\,\Xi$.  For our
values of the \MSb input parameters the zero in the Higgs mass counter
term gets shifted as \mbo{\mu_0 \approx 1.4 \power{16}~\gv
\to \mu'_0\approx 7.7 \power {14}~\gv\,.}
We notice that the SM predicts huge CC at \mbo{\mpl}: $\rho_\phi\simeq
V(\phi) \sim 2.77\,\mpl^4\sim 6.13\power{76}~\gv^4$ exhibiting a very
weakly scale dependence (running couplings) and we are confronted with
the question how to get ride of this huge quasi-constant? An
intriguing structure again solves the puzzle.  The effective CC
counterterm has a zero, which again is a point where renormalized and
bare quantities are in agreement:
\bea
\rho_{\Lambda 0}=\rho_{\Lambda\,} +\frac{\mpl^4}{(16\pi^2)^2}\,X(\mu)
\eea
with $X(\mu)\simeq \frac18\,(2C(\mu)+\lambda(\mu))$ which has a zero
close to the zero of $C(\mu)$ when $2\,C(\mu)=-\lambda(\mu)$. Note
that $C(\mu)=-\lambda(\mu)$ is the shifted Higgs transition point.

Again we find a matching point between low energy and high energy
world: $\rho_{\Lambda 0}=\rho_{\Lambda}$ where
the memory of the quartic Planck scale enhancement gets lost, as it should be
as we know that the low energy phase does not provide access to cutoff
effects.

At the Higgs transition as soon as \mbo{{m'}^2 < 0} for \mbo{\mu <
\mu'_0} the vacuum rearrangement of Higgs potential takes place. As a
result at the minimum $\phi_v$ of the potential we should get
$V(0)+V(\phi_v)\sim \left(0.002~\mbox{eV}\right)^4$ about the observed
value of today's CC. How can this be? Indeed, at the zero of
\mbo{X(\mu)} we have
\mbo{\rho_{\Lambda 0}=\rho_{\Lambda\,}} and one
may expect that like the Higgs boson mass another free SM parameter is
to be fixed by experiment here. One might expect $\rho_{\Lambda}$ to
be naturally small, since the $\lpl^4$ term is nullified at the
matching point. Note that the huge cutoff prefactors act as amplifiers
of small changes in the effective SM couplings. But how small we
should expect the low energy effective CC to be? In fact, in the LEESM
scenario neither the Higgs mass nor the CC are really free parameters
in the low energy world.  The Higgs mass, more precisely the Higgs
self coupling, has to be constrained to a window where the Higgs
potential remains stable up to the Planck scale, and the CC which
triggers inflation gets tuned down by inflation to lie in the ballpark
of the critical density of a flat universe.
\section{Inflation and reheating}
In contrast to standard scenarios of modeling the evolution of the
early universe, SM cosmology is characterized by the fact that almost
everything is known, within uncertainties of the parameters and
perturbative approximations. In LEESM cosmology the form of the
potential is given by the bare SM Higgs potential
$V(\phi)=\frac{m^2}{2}\,\phi^2+\frac{\lambda}{24}\,\phi^4$, the
parameters are known, calculable in terms of the low energy
parameters, the only unknown is the magnitude of the Higgs field. The
latter must be large -- trans-Planckian -- in order to get the
required number of \textit{e-folds} $N$ (expansion factor
$a(t_e)/a(t_i)=\exp H\,(t_e-t_i)=\exp N$, where $a(t)$ is the
Friedmann-Robertson-Walker radius of the universe, $t_i$ the begin of
inflation and $t_e$ the end of inflation and $H$ the Hubble
constant). For our set of \MSb input parameters we require
$\phi_0=\phi(\mu=\mpl) \approx 4.5\,\mpl$. At start the slow-roll
condition $V(\phi) \gg \frac12\,\dot{\phi}^2$ is well satisfied, by the fact
that in the symmetric phase the mass term as well as
$V(0)=\braket{V(\phi)}$ are huge. Because of the large initial field
strength $\phi_0$, however, the interaction term is actually
dominating for a short time after the initial Planck time $\tpl$. The
field equation
\mbo{\ddot{\phi}+3H\dot{\phi}=-V'(\phi)} then predicts a dramatic
decay of the field, \mbo{\phi(t)=\phi_0\,\E^{E_0\,(t-t_0)}} with
\mbo{E_0=\sqrt{2\lambda}/(3\sqrt{3}\ell)\approx
4.3\power{17}~\gv\,,\,\,V_{\mathrm{int}}\gg V_{\mathrm{mass}}} and
shortly after \mbo{E_0=m^2/(3\ell\sqrt{V(0)})\approx
6.6\power{17}~\gv\,,\,\,V_{\mathrm{mass}}\gg V_{\mathrm{int}}}
[\mbo{\ell^2=8\pi G_N/3}], such that in almost no time, still under
slow-roll conditions, the mass term dominates and for what follows the
field equation is a damped harmonic oscillator. The universe thus
undergoes an epoch of Gaussian inflation~\cite{Ade:2013ydc}
before the oscillations set
in. In the symmetric phase the four Higgses are very heavy and rather
unstable. The Higgses decay predominantly (largest Yukawa couplings)
into as yet massless top--anti-top pairs and lighter
fermion--anti-fermion pairs \mbo{H\to t\bar{t},\,b\bar{b},\,\cdots}
thereby reheating the young universe which just had been cooled down
dramatically by inflation.  Preheating is suppressed in SM inflation
as in the symmetric phase bosonic decay channels like $H\to WW$ and
$H\to ZZ$ are absent at tree level. The CP violating decays
\mbo{H^+\to t\bar{d}} [rate \mbo{\propto y_ty_d\,V_{td}}]\,\mbo{H^-\to
b\bar{u}} [rate \mbo{\propto y_by_u\,V_{ub}}] likely are important for
baryogenesis.  After the electroweak phase transition which closely
follows the Higgs transition where $m^2$ in the Higgs potential
changes sign, the now heavy top quarks decay into normal matter as
driven by CKM~\cite{CKM} couplings and phase space. At these scales
the $B+L$ violating dimension 6
operators~\cite{Weinberg:1979sa,Buchmuller:1985jz,Grzadkowski:2010es}
can still play a key role for baryogenesis and together with decays
like \mbo{t\to d e^+\nu} provide CP violating reactions during a phase
out of thermal equilibrium. For details
see~\cite{Jegerlehner:2013cta,Jegerlehner:2013nna,Jegerlehner:2014mua}.
For a very different model of Higgs inflation, which has barely
something in comment with our LEESM scenario,
see~\cite{Bezrukov:2010jz,Bezrukov:2007ep,Bezrukov:2014bra,Bezrukov:2014ipa}.

\section{Remark on Trans-Planckian Higgs fields}
If the SM Higgs is the inflaton, sufficient inflation requires
trans-Planckian magnitude Higgs fields at the Planck scale. At the
cutoff scale the low energy expansion obviously gets obsolete and
likely we cannot seriously argue with field monomials and the operator
hierarchy appearing in the low energy expansion. What is important is
that the field is decaying very fast. Formally, given a truncated
series of operators in the potential, the highest power is dominating
in the trans-Planckian regime. One then expects that for some time
the $\phi^4$ term of the potential is dominating, the decay of the
field is then exponential, for higher dimensional operators it is faster than
exponential, such that the field very rapidly reaches the Planck- and
sub-Planck regime. This means that the mass term is dominating after a very
short period and before the kinetic term becomes relevant and
slow-roll inflation ends. So fears that in low energy effective
scenarios with trans-Planckian fields higher order operators would
mess up things are not in any sense justified\footnote{The
constructive understanding of LEETs we have learned from Ken Wilson's
renormalization semi-group, based on integrating out short distance
fluctuations. This produces all kinds, mostly of irrelevant higher
order interactions. A typical example is the Ising model, which by
itself seen as the basic microscopic system has simple nearest
neighbor interactions only and by the low energy expansion develops a
tower of higher order operators, which at the short distance scale are
simply absent altogether. Such operators don't do any harm at the
intrinsic short distance scale.}. Obviously, without the precise
knowledge of the Planck physics, very close to the Planck scale we
never will be able to make a precise prediction of what is
happening. This however seems not to be a serious obstacle to
quantitatively describe inflation and its properties as far as they
can be accessed by observation. The LEESM scenario in principle
predicts not only the form of the effective potential not far below
the Planck scale but also its parameters and the only quantity not
fixed by low energy physics is the magnitude of the field at the
Planck scale. We also have shown that taking into account the running
of the parameters is mandatory for understanding inflation and
reheating and all that.

Trans-Planckian fields are not unnatural in a low energy effective
scenario as the Planck medium exhibits a high temperature and
temperature fluctuations make higher excitations not improbable.
While the Planck medium will never be accessible to direct
experimental tests, a phenomenological approach to constrain its
effective properties is obviously possible, especially by CMB
data~\cite{Ade:2013uln}.

In the extremely hot Planckian medium, the Hubble constant in the
radiation dominated state with effective number $
g_*(T)=g_B(T)+\frac78\,g_f(T)=102.75$ of relativistic degrees of
freedom is given by $H=\ell \sqrt{\rho}\simeq
1.66\,\left(k_B T\right)^2\sqrt{102.75}$ $\mpl^{-1}$, at Planck time
$ H_i \simeq 16.83\,\mpl$ such that a Higgs field of size
$\phi_i \simeq 4.51\, \mpl$, is not unexpectedly large and could
as well also be bigger.

Often it is argued that trans-Planckian field are unnatural in
particular in a LEET scenario~\cite{Lyth:1996im}. I cannot see any argument against
strong fields and LEET arguments (ordering operators with respect to a
polynomial expansion and their dimension) completely loose their sense
when \mbo{E/\lpl ~\gapprox~ 1}.

Provided the Higgs field decays fast enough towards the end of
inflation we expect the mass term to be dominant such that a Gaussian
fluctuation spectrum prevails. The quasi-constant cosmological constant $V(0)$
at these times mainly enters the Hubble constant $H$ and does not affect
the fluctuation spectrum. 

\section{The self-organized cosmological constant}
In principle, like the Higgs mass in the LEESM, also
\mbo{\rho_{\Lambda\,}} is expected to be a free parameter to
be fixed by experiment. This we would expect in a standard
Einstein-Friedman cosmology not exhibiting a large CC term.  However, the
situation is different if an inflaton exists providing an appropriate
CC. Inflation is fine-tuning the total energy density to unity
$\Omega_{\rm tot}=1$, in units of the critical density defined as a
boarder density between a positively curved universe exhibiting
sufficient mass to stop matter to escape for ever and a negative
curved universe with too little mass to stop the expansion.  Thus
inflation means that the universe after the inflation phase has a
particular energy density $\rho_0=\rho_{EdS}$ of a flat Einstein-de
Sitter universe. This is remarkable as the value is fixed irrespective
of the initial energy density. If there is too much ``mass'' space
adjusts itself such that the same universal density is reached after
the inflation epoch. This also solves the cosmological constant
problem of the SM. The typical problem is that in general one gets a
CC which is way too big and this looks to be a tremendous fine-tuning
problem. For the SM this concerns the contribution to the vacuum
density via the Higgs VEV in the broken phase, as well as the
contributions from spontaneous breakdown of chiral symmetry, which are
much to big and even of wrong sign. However, if inflation is at work,
the final vacuum density is fixed, whatever the initial
density has been. Given that $\Omega_{\rm tot}=\Omega_\Lambda+\Omega_{\rm
DM}+\Omega_{\rm BM}+\Omega_{\rm rad}=1$ with $1>\Omega_{\rm
DM}>\Omega_{\rm
BM}>\Omega_{\rm rad}>0$  we know that $\Omega_\Lambda$ must be of order $\Omega_{\rm
tot}$. As a non-vanishing \mbo{\rho_{\Lambda 0}} is needed
to provide inflation in any case, it is not unlikely that the other
components contributing to the total energy density do not saturate
the bound. This means that the fine-tuning is dynamically enforced by
inflation and the value $\Lambda_{\rm obs}=\Omega_{\Lambda}\times \kappa \simeq 1.6517
\power{-56}~\mathrm{cm}^{-2}$ [$\kappa=\frac{8\pi G_N}{3c^2}$] is
actually not far from the critical total energy density of the flat
universe $\rho_{0,\mathrm{crit}}=\rEdS=\frac{3 H_0^2}{8 \pi G_N}=1.878
\power{-29}\,h^2\,\mathrm{gr/cm}^3$
where $H_0$ is the present Hubble constant, and \mbo{h=0.67\pm0.02} its value in
units of 100 km s$^{-1}$
Mpc$^{-1}$. Today's dark energy density is $\rho_{0\Lambda}=\Omega_\Lambda\,\rho_{0,\mathrm{crit}}$,
with $\Omega_\Lambda=0.74(3)$. 
While $\Omega_{\rm rad}$ and very likely $\Omega_{\rm BM}$ are essentially
SM predictions if we include the $B+L$ violating dimension 6
four-fermion operators, $\Omega_{\rm DM}$ is the only missing piece
which remains an open problem and definitely requires additional
beyond the SM physics. This also concerns contributions from quark and
possible gluon condensates, which we do not explicitly consider here.

\begin{figure}
\centering
\includegraphics[height=4.5cm]{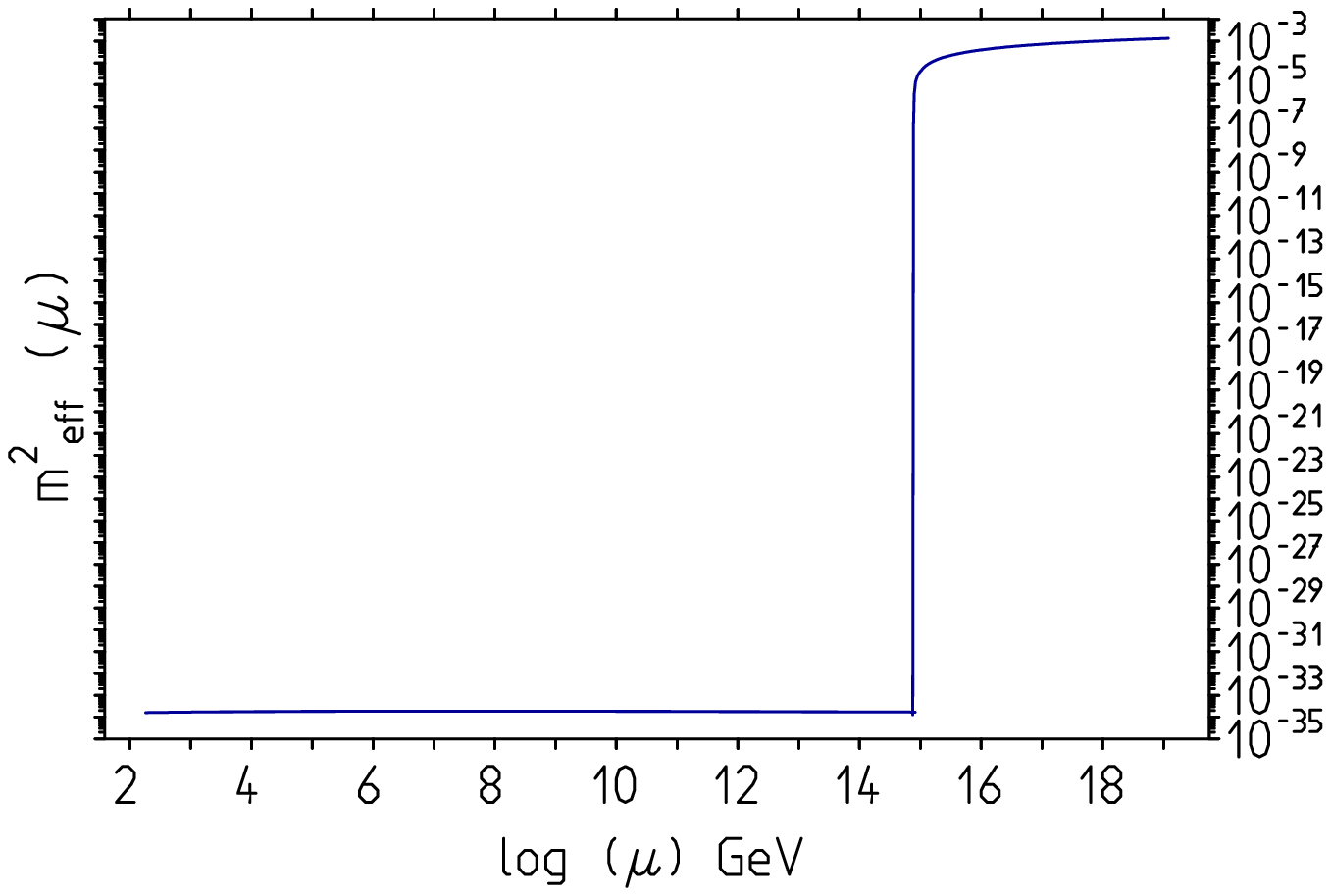}
\includegraphics[height=4.5cm]{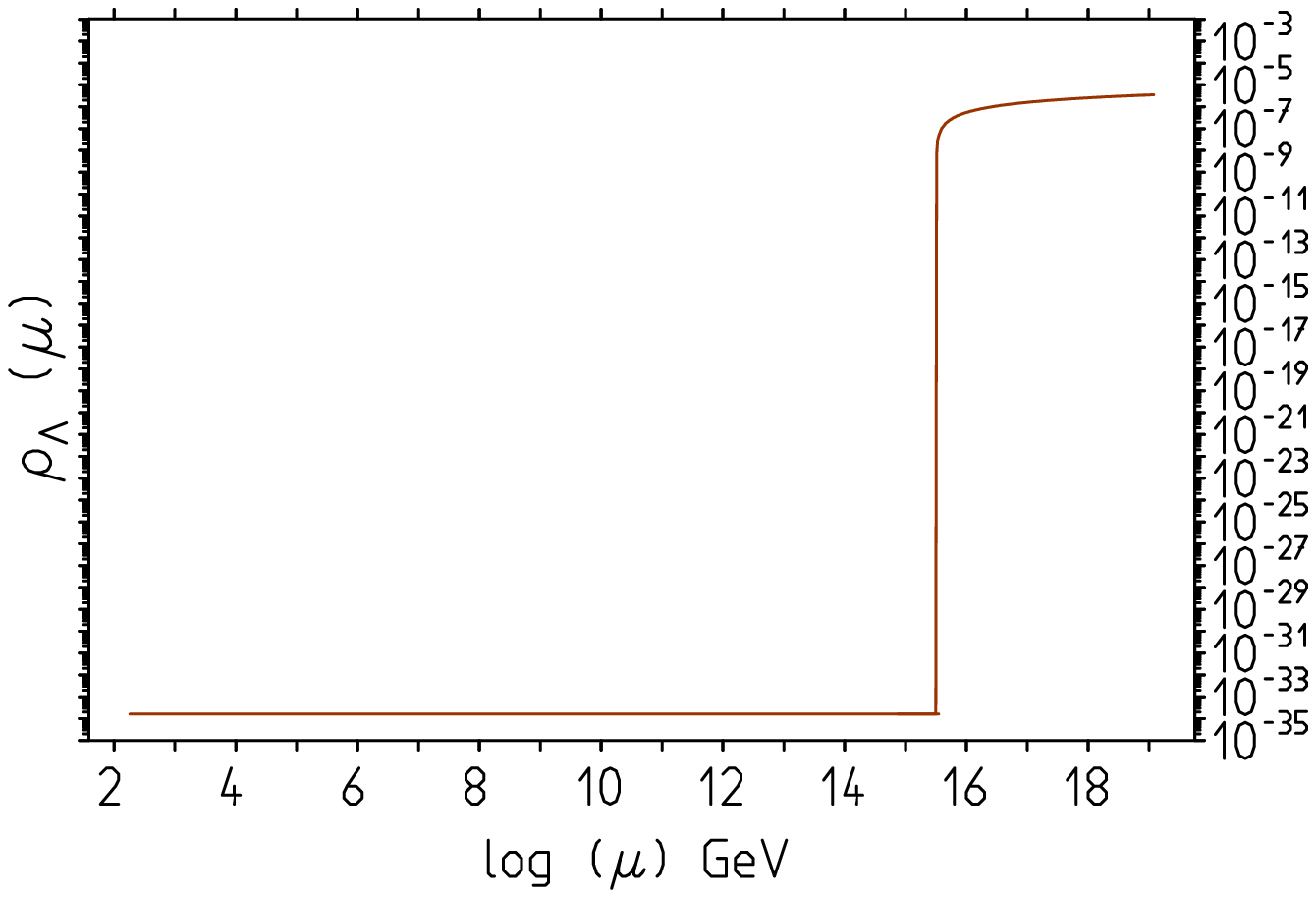}
\caption{The Higgs potential
effective $m^2$ [left] and the dark energy density [right] in Planck
mass units as a function of the energy scale $\mu$ in the SM. Below
the matching point $\mu_{\rm CC}$, where $\rho_\Lambda \simeq 1.6
\power{-47}$ in Plank mass units, we show a scaled up $\rho_\Lambda
\power{13}$ value of the present dark energy density $\mu_\Lambda^4$
with $\mu_\Lambda\simeq 0.002~\mathrm{eV}$.}
\label{fig:CCandm2x} 
\end{figure}

So the solution of the cosmological constant problem is a dynamical
consequence of inflation and overlarge looking values at earlier
epochs in the evolution of the universe just means that the CC is a
time-dependent dynamical quantity. Provided the SM for the specific
conspiring input parameters yields a stable Higgs potential, inflation
and the CC itself are SM ingredients leading to a highly self-consistent
conspiracy which shapes the universe. Fig.~\ref{fig:CCandm2x} shows
the development of the quadratically and the quartically enhanced terms
in the symmetric phase of the SM, and its matching to the
low energy world.
This scenario only seems to be natural if the running of the SM
couplings is not affected by physics beyond the SM in a substantial
way. This LEESM scenario does not exclude the existence of new
physics, however, possible new effects should be natural in the low
energy effective framework. Still axions are a very good candidate to
solve the strong CP problem and eventually provide the missing dark
matter. Similarly, it would be natural that the right-handed singlet
neutrinos are Majorana fermions, which naturally would exhibit a large
Majorana mass and trigger a sea-saw mechanism which would explain the
lightness of the neutrinos. Also an unbroken confining $SU(4)$ sector
could be there forming bound mesons which could constitute the missing
dark matter. Like normal baryonic matter is essentially hadronic
binding energy, bound in fermions, also dark matter could be condensed
energy, bound in bosons. In contrast
a supersymmetric or grand unified extension of the SM would not fit
into the picture. Surprisingly also a fourth family would completely
deteriorate our scenario. I think two points are very much in favor of
a change of the game; Higgs vacuum stability or very close to
stability and why should we need two different scalar field, one for
the Higgs mechanism and one as an inflaton, if one can do what we
need? I agree that it is against any reasonable expectation to believe
that the SM should hold up to the Big Bang. However, fact is that also
cosmological and astrophysical observations have not given any definite
hint for non-SM physics with the big exception of dark matter.

\section{Summary}
The Higgs has two different functions in our world: 1) it has to
render the effective low energy electroweak theory (massive
vector-boson and fermion sector) renormalizable. In the broken low
energy phase the Higgs acquires the vacuum condensate which provides
masses to all massive fields including the Higgs boson itself. Key
point are the many new Higgs-exchange forces necessary to render the
low energy amplitudes renormalizable. 2) in the symmetric phase the
four very heavy Higgses generate a huge dark energy, which causes
inflation. After inflation has ended and we are out of equilibrium the
Higgses are decaying predominantly into the heaviest fermions pairs
which provides the reheating of the inflated universe. The universe
cooling further down then pushes the universe into the Higgs phase,
where the particles acquire their masses. The predominating heavy
quarks decay into the light ones which later form the baryons and
normal matter. This scenario is possible because of the quadratically
enhanced Higgs boson mass and the quartically enhanced dark energy,
which show up in the symmetric phase of the SM before the Higgs
transition. The existence of such relevant operator effects in my
opinion are supported by observation, in particular by observed
inflation patterns, meaning that the hierarchy as well as the
cosmological constant ``problems'' reflect important properties of the
SM needed to understand the evolution of the early universe (for
different opinions
see~\cite{Aoki:2012xs,Blanke:2013uia,Tavares:2013dga,Masina:2013wja,Bian:2013xra}). Consolidation
of our bottom-up path to physics near the Planck scale will sensibly
depend on progress in high precision physics around the EW scale
$v$. Especially, Higgs and a top-pair factories will play a key role
in this context.

A final remark concerning the testability of our Higgs inflation
scenario: any discovery of physics beyond the SM which has its
motivation in the presumed hierarchy problem of the SM, like a super
symmetric extension, extra dimensions etc., would spoil the delicate
balance of SM effective couplings, on which our scenario relies. Also
Grand Unified Theory extensions or even such straight forward
extensions like a fourth fermion family, would likely rule out the
Higgs as an inflaton and as the source of the dark energy.

\bigskip

\noindent
\textbf{Acknowledgments:}\\
I would like to thank Stefan Schaefer for carefully reading the
manuscript and to Mikhail Kalmykov for inspiring discussions.

\bigskip


\end{document}